\documentclass[amsmath,amssymb,aps,prb,citeautoscript,superscriptaddress,twocolumn,showpacs,10pt]{revtex4-1}

\usepackage{graphicx}
\usepackage{dcolumn}
\usepackage{bm}

\usepackage{mathrsfs}

\graphicspath{{figures/}}
\begin{document}

\title{Optimal Control Theory for Time-Dependent Quantum Transport}

\author{Yu Zhang}
\email{zhy@northwestern.edu}
\affiliation{
Center for Bio-inspired Energy Science, Northwestern University, Chicago, Illinois 60611}
\affiliation{
Department of Chemistry, Northwestern University, Evanston, Illinois, 60208, USA
}

\date{\today}

\begin{abstract}
Optical techniques have been employed to coherently control the
quantum transport through nanojunctions. Conventional works
on optical control of quantum transport usually applied a tailored
electrical pulses to perform specific tasks. In this work, an opposite
way is employed and a time-dependent driving field is searched to force
the system behave in desired pattern. In order to achieve the goal, an 
optimal control theory for time-dependent quantum transport is developed. 
The theory provides a theoretical tool for the design of driving field 
to control the transient current through a nano junction along a prescribed
pattern. The optimal control field is searched by minimizing a control
functional. Corresponding equations of motions are derived accordingly
to efficiently search for the optimal control field.
The development of optimal control theory for time-dependent quantum
transport enables the ultra-fast and precise control of current by 
electrical field.
\end{abstract}


\maketitle

\section{Introduction}
Quantum transport in nanojunctions has been of great
research interest~\cite{Kohler2005379,Gaury20141,
PhysRevB.50.5528,nitzan2003}
Previous researches in this field were to measure or calculate
the current-voltage characteristics. In recent years, 
time-resolved studies have attracted more and more attentions~\cite{
RevModPhys.83.349,PhysRevB.71.205304,jcp2938087,zhengjcp2010,PhysRevB.66.205320,
*PhysRevB.69.205316,PhysRevB.75.195127,zhang2014,jz5007143}, 
which give insight on the evolution of  transient current upon a given bias voltage 
or external field. Study of transient transport is of fundamental importance, which helps
us understand whether does a steady-state exist and how is it reached. 
Theoretically, time-dependent approaches have been developed accordingly to uncover
the transient phenomena of quantum transport. These approaches are 
broadly categorized under single-particle approach and many-body theory, 
including quantum Monte-Carlo (QMC)~\cite{RevModPhys.83.349}, quantum 
master equation (QME)~\cite{PhysRevB.71.205304}, hierarchical equation 
of motion (HEOM)~\cite{jcp2938087,zhengjcp2010}, scattering matrix 
approach~\cite{PhysRevB.66.205320,*PhysRevB.69.205316}, non-equilibrium
Green's function (NEGF) and time-dependent density functional theory (TDDFT) based
approaches~\cite{PhysRevB.75.195127,zhang2014,jz5007143,xiejcp2012}, etc.

Despite these theoretical methods differ a lot from each other,
they share one common feature, transient current upon a given
external stimulus, such as bias/gate voltage or external field,
is calculated. However, there is a trend to take a step further
and control the current or charge migration by tailoring stimulus.
Especially, people are trying to tune laser pulse to coherently
control the current by tuning the relative amplitude and phase
of two laser pulses~\cite{nat2013,PhysRevB.87.115201,nature11720}.
Such ideas have also been applied to control
chemical reactions, where chemical reaction are influenced by
femto-second laser pulses such that a specific reaction
gets enhanced or suppressed~\cite{Assion30101998,B305305A,B409446H}.
Consequently, optimal control of quantum systems has attracted more
and more research efforts due to its fundamental importance and potential
applications~\cite{PhysRevB.87.241303,JPB2007,PhysRevE.79.056704,
PhysRevB.77.085324,PhysRevLett.98.157404,njp125028,Levis27042001,
PhysRevB.86.205308,PhysRevLett.109.153603,EPL53001,gross2014}.


Mathematically, optimal control theory is a general idea which has been
used in many problems of classical mechanics. Later it was employed to many other
research fields including quantum dynamics~\cite{PhysRevLett.109.153603,EPL53001,
gross2014,PhysRevB.77.075321,LIGQ2010}. 
In previous works, Kleinekath\"{o}fer and coworkers have combined optimal 
control theory and QME to control quantum transport with tailored 
fields~\cite{PhysRevB.77.075321,LIGQ2010}. However, the many-particle nature
of QME and second-order perturbation treatment of system-lead coupling limit 
its application to small systems and small coupling regime.
In this work, an optimal control theory (OCT) for transient quantum
transport is developed by combining optimal control theory with NEGF 
based time-dependent quantum transport theory.
The article is  organized as follows. Sec.~\ref{sec1} 
introduces time-dependent quantum transport theory. The optimal 
control theory for time-dependent quantum transport is introduced
in details in Sec.~\ref{sec:oct}. Finally, Sec.~\ref{sec:conclusion}
summarizes the manuscript.

\section{Time-dependent quantum transport theory}
\label{sec1}
The system of interest is a device sandwiched by two leads (extension
to multi-lead is formally straightforward). The corresponding 
Hamiltonian reads
\begin{equation}
H=H_S + \sum_\alpha [H_\alpha +H_{\alpha S}],
\end{equation}
where $H_S$ and $H_\alpha$ are the Hamiltonians of the device 
and lead $\alpha$, respectively;
Tight-binding (TB) model Hamiltonians are considered in this work.
$H_{\alpha S}$ is the interaction Hamiltonian between device
and lead $\alpha$. The Hamiltonian of the device region reads
$H_S =\sum_{mn}h_{mn}(t)c^\dag_m c_n$,
where $c^\dag_m$ and $c_n$ are the electronic creation and
annihilation operators in the device region, respectively; 
$h_{mn}(t)$ is the time-dependent TB Fock matrix. In presence
of external field, $h(t)=h_0-d E(t)$,
where $d$ is the dipole matrix. 
The Hamiltonian of lead $\alpha$ is
$H_\alpha= \sum_{k_\alpha}\epsilon_{k_\alpha}(t)c^\dag_{k_\alpha}c_{k_\alpha}$,
where $c^\dag_{k_\alpha}$ and $c_{k_\alpha}$ are the electronic creation
and annihilation operators in the lead $\alpha$, respectively.
$\epsilon_{k_\alpha}(t)$ is the single-particle energy,
the time-dependence of which comes from applied bias voltage or
external field. The variation of single-particle
energy in lead $\alpha$ upon time-dependent bias is assumed to be
$\epsilon_{k_\alpha}(t)=\epsilon^0_{k_\alpha}+\Delta_\alpha(t)$,
with $\Delta_\alpha(t)$ being the voltage applied on lead $\alpha$.
The interaction Hamiltonian between device and lead $\alpha$ reads
$H_{\alpha S} = \sum_{k_\alpha,m}(V_{k_\alpha m}
c^\dag_{k_\alpha}c_m +\text{H.c.})$,
where $V_{k_\alpha m}$ is the coupling strength.

To describe the transient transport, we examine the dynamics
of reduced single-particle density matrix (RSDM).
With the Hamiltonian described above, the equation 
of motion (EOM) of the RSDM reads~\cite{PhysRevB.75.195127}
\begin{equation}\label{eom:densitymatrix}
i\dot{\rho}(t)=[h(t),\rho(t)]-\sum_{\alpha}
[\varphi_\alpha(t) -\varphi^\dag_\alpha(t)],
\end{equation}
the dissipation matrix $\varphi_{\alpha}(t)$  in above equation
denotes the interaction between the device and lead $\alpha$,
which is responsible for the particle dissipation.
Within the framework of NEGF approach, $\varphi_\alpha(t)$ is
\begin{equation}\label{eq:definevarphi}
\varphi_\alpha(t)=i\int^t_{-\infty}d\tau
[G^<(t,\tau)\Sigma^>_{\alpha}(\tau,t)-
G^>(t,\tau)\Sigma^<_\alpha(\tau,t)],
\end{equation}
where $G^<(t,\tau)$ and $G^>(t,\tau)$ are the lesser
and greater Green's function of device, respectively.
$\Sigma^<_\alpha(t,\tau)$ and $\Sigma^>_\alpha(t,\tau)$ are the
lesser and greater self-energies due to
the coupling between device and lead $\alpha$, respectively.
The lesser and greater self-energy can be obtained from the
~\cite{PhysRevB.50.5528}
\begin{equation}\label{eq:selfenergy}
\Sigma^{<,>}_\alpha(\tau,t)=\pm
2i\int \frac{d\epsilon}{2\pi} f^\pm_\alpha(\epsilon)
e^{i\int^t_\tau[\epsilon+\Delta_\alpha(t_1)]dt_1}
\Lambda_\alpha(\epsilon),
\end{equation}
where $f^\pm_\alpha(\epsilon)=1/(e^{\pm \beta(\epsilon-\mu_\alpha)}+1)$ is
the Fermi distribution, with $\beta$ being the inverse temperature.
$\Lambda_\alpha(\epsilon)$ is the line-width function which is
related to the density of state (DOS) of lead and device-lead
coupling strength 
$[\Lambda_\alpha(\epsilon)]_{mn}=
\pi \sum_{k_\alpha}\delta(\epsilon-\epsilon_{k_\alpha})
V^*_{k_\alpha,m}V_{k_\alpha,n}$.

Eq.(\ref{eom:densitymatrix}) and Eq.(\ref{eq:definevarphi})
are the general formalism for open
electronic systems coupled with non-interacting leads.
$\varphi_\alpha(t)$ is corresponding to the net rate of electron
going through the interface between lead $\alpha$ and device. 
The transient current can be evaluated by tracing the 
dissipation matrix $\varphi_\alpha(t)$:
\begin{equation}\label{eq:current}
I_\alpha(t)=i\text{Tr}[\varphi_\alpha(t)-\varphi^\dag_\alpha(t)]
=-2\text{Im}\text{Tr}[\varphi_\alpha(t)].
\end{equation}

The RSDM can be obtained by performing time propagation of 
Eq.(\ref{eom:densitymatrix}). The complexity now lies in the 
evaluation of the dissipation matrix $\varphi_\alpha(t)$. In order
to implement this method to simulate realistic systems from
first-principles, an efficient method to deal with 
$\varphi_\alpha(t)$ is desirable. To achieve this, the WBL 
approximation is employed, which involves the following 
assumptions for the leads:
(i) band widths are assumed to be infinitely large;
(ii) line-widths are assumed to be energy-independent,
i.e., $\Lambda_\alpha(\epsilon)=\Lambda_\alpha$,
where $\Lambda_\alpha=\pi\sum_{k_\alpha}|V|^2\delta(\epsilon_f-
\epsilon_{k_\alpha})$
is the line-width function evaluated at Fermi energy $\epsilon_f$
of the unbiased system. To further improve the calculation 
efficiency, Pad\'{e} expansion approximation is applied to Fermi
distribution function~\cite{hu:244106}. The accuracy of Pad\'{e} 
expansion is determined by the expansion order.
Based on Pad\'{e} expansion and WBL approximation,
the integration in Eq.(\ref{eq:selfenergy}) can be evaluated
analytically through contour integration and residue theorem, the
resulting expression of self-energy is rewritten as
\begin{equation}\label{eq:selfenergypade}
\Sigma^{<,>}_\alpha(\tau,t)\approx
\pm  \frac{i}{2}\delta(t-\tau)\Lambda_\alpha
+x \sum^N_k \Sigma^x_{\alpha k}(\tau,t),
\end{equation}
where $x=sgn(t-\tau)$. The sign $x$ corresponds to upper ($+$) or
lower half plane ($-$) contour integration.
$\Sigma^\pm_{\alpha k}(\tau,t)$ is defined as
$
\Sigma^\pm_{\alpha k}(\tau,t)=\frac{2}{\beta}\eta_k
e^{i\int^t_\tau\epsilon^{\pm}_{\alpha k}(t_1)dt_1}\Lambda_\alpha,
$
where $\epsilon^\pm_{\alpha k}(t)=\pm i\zeta_k/\beta+\mu_\alpha+\Delta_\alpha(t)$.
$\pm i\zeta_k/\beta+\mu_\alpha$ are the $k$th Pad\'{e} poles in the
upper and lower half plane, respectively; $\eta_k/\beta$ is the
corresponding coefficient. Consequently, the dissipation
matrix is rewritten as
\begin{equation}\label{eq:varphialpha}
\varphi_\alpha(t)
=i[\rho(t)-1/2]\Lambda_\alpha+\sum^N_k \varphi_{\alpha k}(t).
\end{equation}
where $\varphi_{\alpha k}(t)$ is the component of the dissipation
matrix, which is evaluated through its EOM. Within the WBL approximation, the second term 
on the RHS of above equation is written as
\begin{eqnarray}\label{eq:varphik}
\varphi_{\alpha k}(\tau)=
-\frac{2\eta_k}{\beta}\int^\tau_{-\infty}dt_1
\tilde{U}_{\alpha k}(\tau,t_1)\Lambda_\alpha
\end{eqnarray}
where $\tilde{U}_{\alpha k}(\tau,t_1)$ can be regarded as the
propagator, which is defined as
\begin{equation}
\tilde{U}_{\alpha k}(\tau,t_1)=e^{i\int^\tau_{t_1}
[\epsilon_{\alpha k}(t')-h(t')+i\Lambda]dt'}.
\end{equation}
It is obvious that $\varphi_{\alpha k}(t)$ can be easily 
calculated from its EOM~\cite{PhysRevB.87.085110}.

\section{Optimal control theory for time-dependent quantum transport}
\label{sec:oct}
The key ingredient of optimal control is to determine  the
electrical field that can lead to a predefined effect on the current 
through junctions. This is achieved by using the optimal 
control theory (OCT) which optimizes a control functional.
Considering the optimal control of current, the goal is to search for an optimal
control field such that the current follows a target pattern as
best as possible. Mathematically, the different between the 
desired current pattern $P(t)$ and the current obtained from
the calculation at each iteration is to be minimized. 
Hence, we can define a control functional as
\begin{equation}
J_I[E]=\int^{t_f}_{t_0} dt \left[P(t)-I(t)\right]^2,
\end{equation}
where $I(t)$ is the time-dependent current calculated by the
method mentioned in previous section. Obviously, an electrical 
field that makes $J_I[E(t)]=0$ is the optimal control field.
To ensure the convergence, an additional part is added to the 
control functional,
\begin{equation}\label{eq:controlf}
J[E]=J_I[E]+\frac{\lambda}{2}\int^{t_f}_{t_0}\frac{[E(t)-\tilde{E}(t)]^2}{s(t)}
\end{equation}
with $\tilde{E}(t)$ being the electrical field of the previous
iteration step. The penalty parameter $\lambda$ is a Lagrange
multiplier and a time-dependent function $s(t)$ is introduced
to avoid sudden switch-on and switch-off behavior of the control
field. For the optimization of the function as described in 
Eq.(\ref{eq:controlf}), the functional derivative with respect
to the field $E(t)$ should vanish, i.e.,
$\frac{\delta J[E]}{\delta E(t)}=0$. 
This yields the condition for external fields,
\begin{equation}
E(t)=\tilde{E}(t)-\frac{s(t)}{\lambda}\frac{\delta J_I[E]}{\delta E(t)}.
\end{equation}
Therefore, once the fractional derivative of $J_I(E)$ with
respect to external field is known, the optimal control
field is obtained. According to the definition of $J_I[E]$,
$\frac{\delta J_I[E]}{\delta E(t)}$ depends on the functional
derivative of current, i.e., $\frac{\delta I(\tau)}{\delta E(t)}$.
Within NEGF formalism, current is is in terms of self-energies
and Green's functions as given by Eq.(\ref{eq:current}).
Hence, the functional derivative of $J_I[E]$ with respect to 
the external field is written as
\begin{eqnarray}
\frac{\delta J_I[E]}{\delta E(t)}
=-4\int^{t_f}_{t_0}d\tau
[I(\tau)-P(\tau)]\text{Im}\text{Tr}
\left[\frac{\delta \varphi_\alpha(\tau)}{\delta E(t)}\right].
\end{eqnarray}
Because $\varphi_{\alpha}(t)$ is decomposed due to Pad\'{e}
approximation as shown in Eq.(\ref{eq:varphialpha}), the 
functional derivative of $\varphi_\alpha(\tau)$ with respect
to field $E(t)$ can be decomposed accordingly,
\begin{equation}
\frac{\delta \varphi_\alpha(\tau)}{\delta E(t)}=
i\frac{\delta \rho(\tau)}{\delta E(t)}\Lambda_\alpha
+\sum^N_k \frac{\delta \varphi_{\alpha k}(\tau)}{\delta E(t)}.
\end{equation}
The first part on the right hand side (RHS) of above equation
requires the detailed knowledge of density matrix within the
NEGF formalism. While the second part on the RHS of above
equation depends on the propagator as indicated by
Eq.(\ref{eq:varphik}). The analytical forms of the functional
derivatives of $\rho(\tau)$ and  $\varphi_{\alpha k}(\tau)$ with 
respect to $E(t)$ are derived as follows.

According to the definition of $\varphi_{\alpha k}(t)$ as shown
by Eq.(\ref{eq:varphik}), only functional derivative of 
propagator is needed. Because the single-particle energy 
$\epsilon_{\alpha k}$ and Fock matrix are dependent on 
external field, functional derivative of propagator with respect 
to external field is 
\begin{equation}
\frac{\delta \tilde{U}_{\alpha k}(\tau,t_1)}{\delta E(t)}=
-i\vartheta(\tau-t)\vartheta(t-t_1)
\tilde{U}_{\alpha k}(\tau,t)\tilde{d} \tilde{U}_{\alpha k}(t,t_1),
\end{equation}
where $\tilde{d}$ is defined as
$\tilde{d}=\delta[h(t)-\epsilon_{\alpha k}(t)]/\delta E(t)$,
then the functional derivative of $\varphi_{\alpha k}(\tau)$ with
respect to field $E(t)$ is expressed as
\begin{equation}
\frac{\delta \varphi_{\alpha k}(\tau)}{\delta E(t)}=
-i\vartheta(\tau-t)\tilde{U}_{\alpha k}(\tau,t)\tilde{d}
\varphi_{\alpha k}(t).
\end{equation}
Thus, the functional derivative of $\varphi_{\alpha k}(\tau)$
with respect to $E(t)$ is analytically obtained. It should be noticed that
if $\epsilon_{\alpha k}(t)$ is not dependent on $E(t)$, then
$\tilde{d}=\frac{\delta h(t)}{\delta E(t)}=d$.

Now, the problem existed in the combination between NEGF-WBL
and OCT is the how to get the functional derivative of density matrix
$\rho(\tau)$ with respect to external field $E(t)$, i.e.,
$\frac{\delta \rho(\tau)}{\delta E(t)}$.
Within NEGF formalism, density matrix can be evaluated through lesser
Green's function $\rho(t)=-iG^<(t,t)$, while
the letter can be expressed in terms of retarded/advanced Green's
functions and self-energies,
\begin{equation}
G^<(t,t)=\int dt_1\int dt_2
G^r(t,t_1)\Sigma^<(t_1,t_2)G^a(t_2,t).
\end{equation}
Where $\Sigma^<(t_1,t_2)=\sum_\alpha \Sigma_\alpha(t_1,t_2)$
is the total self-energy.
Within WBL approximation, the retarded Green's function
$G^r(t,t_1)$ can be expressed as
\begin{eqnarray}
G^r(t,t_1)=&&-i\vartheta(t-t_1)U(t,t_1),
\end{eqnarray}
where $U(t,t_1)=e^{-i\int^t_{t_1}[h(t')-i\Lambda]dt']}$ is defined
with $\Lambda=\sum_\alpha \Lambda_\alpha$ being
the total line-width function. With the Pad\'{e} approximation
to the Fermi distribution function and WBL approximation, the 
lesser self-energy $\Sigma_\alpha(t_1,t_2)$ is given by 
Eq.(\ref{eq:selfenergypade}).
Hence, the density matrix is rewritten as
\begin{eqnarray}\label{eq:rho}
&&\rho(\tau)=\int dt_1 \sum_\alpha
G^r(\tau,t_1)\Lambda_\alpha G^a(t_1,\tau)
\nonumber\\&&
-i\int dt_1\int dt_2 \sum_{\alpha k}G^r(\tau,t_1)
\Sigma^{x}_{\alpha k}(t_1,t_2)G^a(t_2,\tau).
\end{eqnarray}
Thus, after lengthy derivation, the functional derivative of 
$\rho(\tau)$ with respect to $E(t)$ is
\begin{eqnarray}
\frac{\delta \rho(\tau)}{\delta E(t)}=&&
-i\vartheta(\tau-t)\Big\{U(\tau,t) \left[d, \rho(t)\right]U^\dag(\tau,t)
+\nonumber\\&&
\Big[ U(\tau,t)\tilde{d}\sum_{\alpha k}\varphi_{\alpha k}(t)
\phi_{\alpha k}(t,\tau)+h.c.\Big]\Big\}.
\end{eqnarray}
where $\phi_{\alpha k}(t,\tau)$ is defined as
$\phi_{\alpha k}(t,\tau)=\int_t dt_2\Sigma^x_{\alpha k}(t,t_2)G^a(t_2,\tau)$.
Hence, $\frac{\delta \varphi_\alpha(\tau)}{\delta E(t)}$ can be
readily evaluated. 

Therefore, the functional derivative of $J_I[E]$ with respect to 
external field $E(t)$ is
\begin{eqnarray}
\frac{\delta J_I[E]}{\delta E(t)}=&&
-4\int^{t_f}_{t} d\tau O(\tau)\text{Im}\text{Tr}
\Big\{U^\dag(\tau,t)\Lambda_\alpha U(\tau,t)\left[d,\rho(t)\right]
\nonumber\\&&
+\sum_{\beta k}[\phi_{\beta k}(t,\tau)\Lambda_\alpha
U(\tau,t)\tilde{d}\varphi_{\beta k}(t)+h.c.]
\nonumber\\&&
-i\sum_k\tilde{U}_{\alpha k}(\tau,t)
\tilde{d}\varphi_{\alpha k}(t)\Big\}.
\end{eqnarray}
Where $O(\tau)=I(\tau)-P(\tau)$. Now, the functional
derivative of $J_I[E]$ with respect to external field
is analytically obtained. The difficulty of evaluating
$\frac{\delta J_I[E]}{\delta E(t)}$ is the time-integration.
Defining
\begin{eqnarray}
\chi_\alpha(t)&&=\int^{t_f}_{t}d\tau O(\tau)
U^\dag(\tau,t)\Lambda_\alpha U(\tau,t)
\nonumber\\
\Theta_{\alpha k}(t)&&=i\int^{t_f}_{t}d\tau O(\tau)
\tilde{U}_{\alpha k}(\tau,t)
\nonumber\\
\Upsilon_{\beta k}(t)&&=\int^{t_f}_{t}d\tau O(\tau)
\phi_{\beta k}(t,\tau)\Lambda_\alpha U(\tau,t)
\end{eqnarray}
the functional derivative of current with
respect to external field, $\frac{\delta J_I[E]}{\delta E(t)}$, 
can be rewritten as
\begin{eqnarray}
\frac{\delta J_I[E]}{\delta E(t)}=&&-4\text{Im}\text{Tr}
\Bigg\{\chi_\alpha(t)[d,\rho(t)]
-\sum_{k}\Theta_{\alpha k}(t)\tilde{d}\varphi_{\alpha k}(t)
\nonumber\\&&
+\sum_{\beta k}\left[\Upsilon_{\beta k}(t)\tilde{d}\varphi_{\beta k}(t)+h.c.\right]
\Bigg\}.
\end{eqnarray}
Hence, analytical expression of $\frac{\delta J_I[E]}{\delta E(t)}$
is obtained. Density matrix $\rho(t)$ and $\varphi_{\alpha k}(t)$
can be evaluated via the time-propagation of their EOMs. Only $\chi_\alpha(t)$,
$\Theta_{\alpha k}(t)$ and $\Upsilon_{\beta k}(t)$ remain unknown.

According to the definition of $\chi_\alpha(t)$, 
$\Theta_{\alpha k}(t)$ and $\Upsilon_{\beta k}(t)$, 
they can be evaluated through their EOMs,
\begin{eqnarray}\label{eq:eom2}
\dot{\Theta}_{\alpha k}(t)=&&
-i O(t)-i\Theta_{\alpha k}(t)[\epsilon_{\alpha k}(t)-h(t)+i\Lambda]
\nonumber\\
\dot{\Upsilon}_{\beta k}(t)=&&
-i\frac{2\eta_k}{\beta}\Lambda_\beta\chi_\alpha(t)
+i\Upsilon_{\beta k}(t)[\epsilon_{\alpha k}(t)+h(t)-i\Lambda]
\nonumber\\
\dot{\chi}_\alpha(t)=&&-O(t)\Lambda_\alpha-i\mathcal{R}(t)\chi_\alpha(t).
\end{eqnarray}
where $\mathcal{R}(t)\equiv[h(t)+i\Lambda,\cdot]_+$ and
$[A,B]_+=AB-BA^\dag$. It is obvious that the boundary 
conditions for $\chi_\alpha(t)$, $\Theta_{\alpha k}(t)$ and 
$\Upsilon_{\beta k}(t)$ are at time $t=t_f$ where all the three
quantities are zero. Therefore, EOMs of Eq.(\ref{eq:eom2}) have to be
propagated backwards from $t_f$ to $t_0$ in order to update the
control field. In the contrast, the density matrix $\rho(t)$ is 
propagated forward in time. Eqs.(\ref{eom:densitymatrix}) and (\ref{eq:eom2})
have to be solved iteratively. In short, the numerical procedure
is summarized as follows,
\begin{enumerate}
\item Time-propagation of density matrix and dissipation matrices from $t_0$ to $t_f$ with initial guess of external field;
\item Backward propagation of $\chi_\alpha(t)$, $\Theta_{\alpha k}(t)$ and $\Upsilon_{\beta k}(t)$ from $t_f$ to $t_0$ and update the control field; 
\item Procedure (1) and (2) can repeated iteratively until convergence is achieved. 
\end{enumerate}

However, the procedure (2) requires the storage of density matrix
$\rho(t)$ and the component of dissipation matrix $\varphi_{\alpha k}(t)$
at each time step, which requires large amount of memory.
To reduce the memory requirement, cubic spline interpolation
of control field is employed. The electrical field, $E(t)$, is
approximated by cubic splines with $N+1$ equidistant nodes at 
$t_k=\frac{k}{N}T, k={0,\cdots,N}$. Accordingly, only $N+1$ snapshots
of each quantity at time $t_k, k={0,\cdots,N}$ are needed. Compared 
to the number of time-step used in the propagation (typical simulation
requires thousands of time-steps or tens of thousand time-steps),
spline interpolation can efficiently reduce the memory requirement
significantly.

\section{Summary}
\label{sec:conclusion}
In summary, an optimal control theory for time-dependent quantum
transport is developed. The method combines NEGF-WBL method with
optimal control theory to achieve the optimal control of transient
current. Numerical implementation of present method is also briefly
introduced. By employing the method, the optimal control pulse can be
found upon desired current pattern. The method can also be employed
to find the optimal performance of nano devices, such as optimize 
cooper pair splitting efficiency, photovoltaics, thermoelectric, etc.
Moreover, the approach is expected to be useful in the control of 
other observables in quantum transport.

\begin{acknowledgments}
This work was supported as part of the Center for
Bio-Inspired Energy Science, an Energy Frontier Research Center
funded by the U.S. Department of Energy, Office of Science,
Basic Energy Sciences under Award Number DE-SC0000989-002.
\end{acknowledgments}

\bibliography{ref}

\end{document}